\documentclass{article}
\usepackage{makecell}
\usepackage{spconf,amsmath,graphicx,hyperref}
\usepackage{booktabs}
\usepackage{multirow}
\usepackage{siunitx} 
\usepackage{tabularx} 
\usepackage{subcaption}     
\usepackage{amssymb}        

\title{VoCodec: An Efficient Lightweight Low-Bitrate Speech Codec}
\name{Leyan Yang$^{1,2,\dagger}$, Ronghui Hu$^{1,2,\dagger}$, Yang Xu$^{1,2,\dagger}$, Jing Lu$^{1,2,*}$}
\address{
    $^{1}$Key Laboratory of Modern Acoustics, Nanjing University, Nanjing 210093, Jiangsu, China\\
    $^{2}$NJU-Horizon Intelligent Audio Lab, Horizon Robotics, Beijing 100094, China
}
\begin{document}
\maketitle
\begingroup
\renewcommand\thefootnote{\fnsymbol{footnote}}
\footnotetext[2]{Equal contribution.}
\footnotetext[1]{Corresponding author: \texttt{lujing@nju.edu.cn}}
\endgroup
\begin{abstract}
Recent advancements in end-to-end neural speech codecs enable compressing audio at extremely low bitrates while maintaining high-fidelity reconstruction. Meanwhile, low computational complexity and low latency are crucial for real-time communication. In this paper, we propose VoCodec, a speech codec model featuring a computational complexity of only 349.29M multiply-accumulate operations per second (MACs/s) and a latency of 30 ms. With the competitive vocoder Vocos as its backbone, the proposed model ranked fourth on Track 1 in the 2025 LRAC Challenge and achieved the highest subjective evaluation score (MUSHRA) on the clean speech test set. Additionally, we cascade a lightweight neural network at the front end to extend its capability of speech enhancement. Experimental results demonstrate that the two systems achieve competitive performance across multiple evaluation metrics. Speech samples can be found at \href{https://acceleration123.github.io/}{https://acceleration123.github.io/}.
\end{abstract} 
\vspace{-0.5em} 
\begin{keywords}
speech codec, low computational complexity, low bitrate, generative adversarial network, speech enhancement
\end{keywords}
\vspace{-1.0em} 
\section{Introduction}
\label{sec:intro}

Speech codecs serve the purpose of compressing and decompressing speech signals to enable effective transmission and storage. Traditional codec methods \cite{rfc6716, Dietz2015OverviewOT} rely on signal processing techniques, while also leveraging knowledge of psychoacoustics and speech synthesis to enhance their coding efficiency. Although traditional codecs are efficient for improving perceptual quality, their design requires significant manual effort, including parameter tuning and subjective listening tests.


Neural codecs \cite{DBLP:journals/corr/abs-2502-06490} usually consist of an encoder, a decoder, and a quantizer module. The encoder compresses the speech into a latent representation, while the decoder reconstructs the waveform from the quantized vectors. The quantizer is usually parameterized and jointly trained with the encoder and decoder in an end-to-end manner. Compared to traditional codecs, neural codecs use discrete tokens for compression and reconstruction rather than continuous embeddings. Most neural codecs are based on the VQ-GAN architecture \cite{esser2021taming} and employ diverse discriminators \cite{kumar2020melgan, 2020HiFi, Jang2021UnivNetAN} to improve the perceptual quality of the reconstructed speech. Existing neural codecs have demonstrated remarkable reconstruction quality \cite{soundstream, defossez2022highfi, kumar2023highfidelity}. Beyond the perceptual quality of the reconstructed speech, the low bitrate of neural codecs has long been a research focus in the field, since it is critical for real-time communication. BigCodec \cite{xin2024bigcodec} scales up the model size to compensate for the performance degradation caused by its low bitrate, ultimately achieving high-quality speech reconstruction at a bitrate of 1.04 kbps. WavTokenizer \cite{ji2024wavtokenizer} carefully investigates codebook utilization for speech and select a larger single codebook, attaining a bitrate as low as 0.98 kbps.

Although recent speech codec models have achieved significant progress in recovering high-quality speech at low bitrates, existing models with excellent performance often suffer from two critical drawbacks: high computational complexity and non-causality, rendering them unsuitable for real-time communication\cite{kumar2023highfidelity, xin2024bigcodec}. In addition, noise and reverberation in real-world scenarios may cause severe interference to real-time communication.
The 2025 Low-Resource Audio Codec (LRAC) challenge \cite{lrac} focuses on speech codec under joint resource constraints, which includes
computational complexity, bitrate, and latency. Track 1 targets high reconstruction quality, while Track 2 further addresses coding under noisy and reverberant conditions.

In this paper, we propose VoCodec, a model that simultaneously features low computational complexity, low latency, and support for low-bitrate transmission. VoCodec's encoder and decoder are based on Vocos \cite{siuzdak2023vocos}. To reduce computational overhead and latency, we operate speech codec directly in the time-frequency domain. The final model requires a computational complexity of only 349.29M MACs/s, with
the receiver-side computation accounting for only 144.82M MACs/s. Additionally, UL-UNAS \cite{rong2025ulunasultralightweightunetsrealtime}, a lightweight model for speech enhancement is cascaded at the front end of VoCodec to equip our system with capabilities of noise reduction and dereverberation.

We train the two systems on a large-scale dataset provided by the challenge organizers. Experimental results demonstrate that compared to the baseline models, our methods achieve competitive performance across multiple evaluation metrics.


\begin{figure*}[t!]

\begin{minipage}[b]{1.0\linewidth}
  \centering
  \centerline{\includegraphics[width=17.0cm]{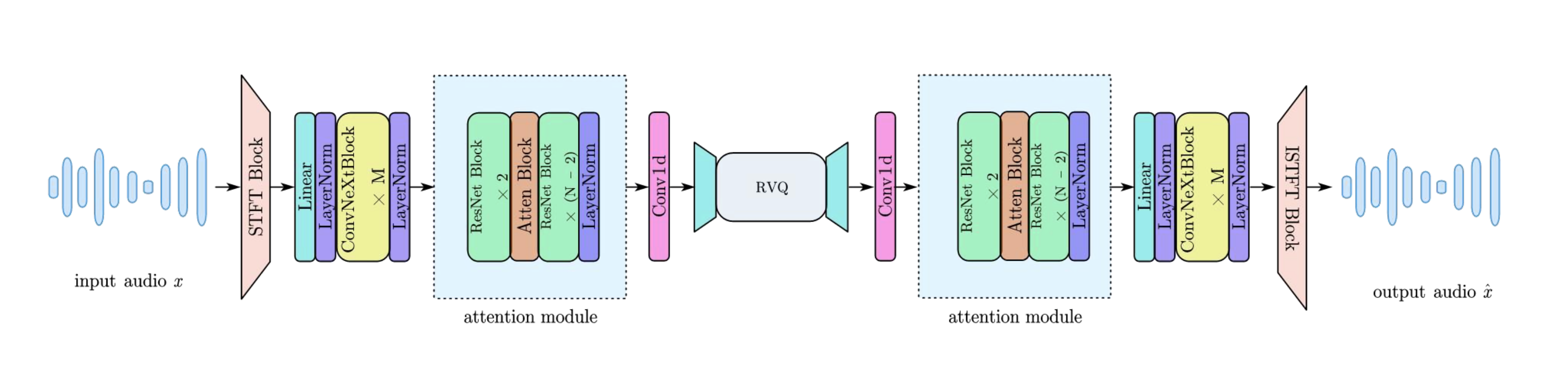}}
  \vspace{-0.8cm}
\end{minipage}


\caption{The architecture of the proposed model}
\label{fig:res}
\end{figure*}

\vspace{-1.0em}

\section{Proposed Model}
\label{sec:codec}

\vspace{-1.0em}

\subsection{Generator}
\label{ssec:arch}

\vspace{-1.0em}

Unlike mainstream time-domain codec models \cite{defossez2022highfi, xin2024bigcodec, ji2024wavtokenizer}, we choose to operate speech codec in the time-frequency domain. The reasons are twofold: 1) Speech exhibits a highly pronounced harmonic structure in the time-frequency domain. 2) Downsampling and upsampling operations of the time dimension can be accomplished by Short-Time Fourier Transform (STFT) and its inverse transform in a single step, thereby reducing computational complexity and latency. By outputting complex spectral coefficients to reconstruct speech, Vocos \cite{siuzdak2023vocos} has achieved state-of-the-art (SOTA) performance in the field of speech synthesis. Inspired by this, we adopt Vocos as the backbone of VoCodec's encoder and decoder. 

The architecture of the proposed VoCodec is depicted in Figure \ref{fig:res}. Given a speech signal $x \in \mathbb{R}^L$, it is first transformed into the complex spectrum $X \in \mathbb{C}^{F \times T}$ via STFT, where $F$ and $T$ denote the frequency and frame axes, respectively. Then the logarithmic magnitude and phase of the complex spectrum $X$ are extracted and concatenated along the frequency axis as the input feature $Z_{in}$, formulated as

\vspace{-1.0em}

\begin{equation}
Z_{in} = \text{Concat}\left( \log(\lvert X \rvert), \mathrm{angle}(X_i, X_r) \right) \in \mathbb{R}^{2F \times T} \tag{1}
\end{equation}

\noindent where $X_r$, $X_i$ denote the real and imaginary parts of the complex spectrum, respectively. The \( |\cdot| \) operator represents the norm of a complex value and Concat($\cdot$) is the concatenation operation. To reduce computational complexity, a fully connected layer is employed to project $Z_{in}$ into a low-dimensional space.

The subsequent encoder follows the improved Vocos architecture in WavTokenizer \cite{ji2024wavtokenizer}, which consists of $M$ stacked ConvNeXt blocks \cite{liu2022convnet} and an attention module. Each ConvNeXt block consists of a depthwise convolution, followed by an inverted bottleneck that projects features into a higher dimensionality and then projects them back using two pointwise convolutions. The attention module aims to improve the VoCodec’s sequence modeling ability: we incorporate $N$ basic ResNet blocks \cite{7780459} and add a self-attention block \cite{NIPS2017_3f5ee243} after the second ResNet block. Meanwhile, to ensure causality and avoid introducing extra latency into the codec model, all convolutions use causal zero-padding with stride 1, and the self-attention layer employs masked attention to guarantee that the model does not access information from future frames.

VoCodec’s quantizer uses the Residual Vector Quantization (RVQ) strategy \cite{1171604}. Following the improved RVQ proposed in DAC \cite{kumar2023highfidelity}, factorized codes and L2-normalization are employed. Factorization performs code lookup in a low-dimensional space (8-D in our model) through a fully connected layer. The L2-normalization of the encoded and codebook vectors converts euclidean distance to cosine distance. The quantizer of our model is applied with 6 layers, each containing 1024 codewords. With an encoder frame rate of 100 Hz, this corresponds to 1 kbps per layer, and 6 kbps in total.

For decoder, it is almost a mirror-symmetric structure of the encoder. However, since we do not perform the upsampling operation in the network, transposed convolutions are not used in the decoder. Additionally, to constrain receiver-side's computational complexity below 150M MACs/s, we remove the inverted design in the ConvNext block and the group convolution is used in the ResNet block. The network ultimately generates complex spectral coefficients and the speech is reconstructed via the inverse STFT \cite{siuzdak2023vocos}.

\vspace{-1.0em}
\subsection{Discriminator}
\vspace{-1.0em}

Since we operate speech codec in the time-frequency domain, intuitively, the multi-scale STFT discriminator \cite{defossez2022highfi} can further improve the quality of the output audio. A set of window lengths [128, 256, 512, 1024, 2048] is used, and the hop length is fixed to the window length / 4. Moreover, only this discriminator is employed throughout the training process, while other types such as multi-scale discriminator \cite{kumar2020melgan} and multi-period discriminator \cite{2020HiFi} are not used.

\vspace{-1.0em}

\subsection{Combined enhancement and compression}
\label{se}

\vspace{-1.0em}

As the proposed VoCodec directly operates speech codec in the time-frequency domain, a lightweight speech enhancement model based on time-frequency domain masking can be integrated at the front end of the codec model to eliminate the interference of noise and reverberation. Specifically, we cascade the UL-UNAS \cite{rong2025ulunasultralightweightunetsrealtime} and VoCodec models: the speech signal $x$ first passes through UL-UNAS to get the enhanced spectrum $X_{enh}$. Then $X_{enh}$ is preprocessed as mentioned in section \ref{ssec:arch} and fed into VoCodec. Considering that the speech enhancement network introduces certain nonlinear distortions, we freeze the parameters of UL-UNAS and conduct fine-tuning of VoCodec after independently training each one. 

\vspace{-1.0em}

\subsection{Loss functions}
\label{ssec:loss}

\vspace{-1.0em}

When training UL-UNAS, we apply the negative scale invariant SNR (SI-SNR) \cite{8683855} loss $\mathcal{L}_{\text{SI-SNR}}$, the power-compressed spectrum losses $\mathcal{L}_{\text{mag}}$, and $\mathcal{L}_{\text{real/imag}}$ as the loss functions, formulated as

\vspace{-1.0em}

\begin{equation}
\mathcal{L}_{\text{SI-SNR}}(\hat{x}, x) = -\log_{10}\left( \frac{\|\hat{x}_t\|_{2}^{2}}{\|\hat{x} - \hat{x}_t\|_{2}^{2}} \right); \hat{x}_t = \frac{\langle \hat{x}, x \rangle x}{\|x\|_{2}^{2}}
\tag{2}
\label{eq:si_snr}
\end{equation}

\begin{equation}
\mathcal{L}_{\text{mag}}(\hat{X}, X) = \left \| |\hat{X}|^{0.3} - |X|^{0.3} \right \| _{2}^{2}
\tag{3}
\label{eq:mag_loss}
\end{equation}

\begin{equation}
\mathcal{L}_{\text{real/imag}}(\hat{X}, X) = 
\left \| \frac{\hat{X}_{\text{r/i}}}{|\hat{X}|^{0.7}} - \frac{X_{\text{r/i}}}{|X|^{0.7}} \right \| _{2}^{2}
\tag{4}
\label{eq:real_imag_loss}
\end{equation}

\noindent where $x$ and $\hat{x}$ represent clean and enhanced speech, $X$ and $\hat{X}$ are their corresponding spectrograms, the subscripts $\text{r}$, $\text{i}$ represent the real and imaginary parts of the spectrograms, respectively, and $\langle \cdot, \cdot \rangle$ denotes the inner product operator. The general loss function $\mathcal{L}_{se}$ for model training is given by

\begin{equation}
\begin{aligned}
\mathcal{L}_{se} &= \lambda_1 \mathcal{L}_{\text{SI-SNR}}(\hat{x}, x) + \lambda_2 \mathcal{L}_{\text{mag}}(\hat{X}, X) \\
&\quad + \lambda_3 \left( \mathcal{L}_{\text{real}}(\hat{X}, X) + \mathcal{L}_{\text{imag}}(\hat{X}, X) \right)
\end{aligned}
\tag{5}
\label{eq:total_loss}
\end{equation}

\noindent where $\lambda_1$, $\lambda_2$ and $\lambda_3$ are the empirical weights.

For VoCodec, the generator loss $\mathcal{L}_{generator}$ comprises three components: the multi-scale mel-spectrogram loss \cite{kumar2023highfidelity} as the reconstruction loss $\mathcal{L}_{rec}$, the adversarial loss $\mathcal{L}_{g}$ with the feature matching loss $\mathcal{L}_{feat}$ involved, the same codebook $\mathcal{L}_{code}$ and commitment loss $\mathcal{L}_{c}$ in VQ-VAE \cite{NIPS2017_7a98af17} for codebook updates. The discriminator is trained separately with the adversarial loss $\mathcal{L}_{d}$. Formally,

\begin{equation}
    \mathcal{L}_{rec} =  \left \| \log(\mathcal{M} \left( x \right)) - \log(\mathcal{M} \left( \hat{x}\right))  \right \| _{1}
    \tag{6}
    \label{1}
\end{equation} 
\begin{equation}
    \mathcal{L}_{g} = \left \| 1 - D(\hat{x} ) \right \| _{2}^{2}
    \tag{7}
    \label{2}
\end{equation}
\begin{equation}
    \mathcal{L}_{feat} = 2\sum_{l}^{}\left \| D^{l}(x)-D^{l}(\hat{x} ) \right \| _1  
    \tag{8}
    \label{3}
\end{equation}
\begin{equation}
\begin{split}
    \mathcal{L}_{\text{generator}} = & \lambda_{\text{rec}}\mathcal{L}_{\text{rec}} + \lambda_{g}\mathcal{L}_{g} + \lambda_{\text{feat}}\mathcal{L}_{\text{feat}} \\
    & + \lambda_{\text{code}}\underbrace{
        \left\| \text{sg}[\mathbf{z}_e] - \mathbf{e}_k \right\|_{2}^2
    }_{\mathcal{L}_{\text{code}}} + \lambda_{c}\underbrace{
        \left\| \mathbf{z}_e - \text{sg}[\mathbf{e}_k] \right\|_{2}^2
    }_{\mathcal{L}_{c}} 
\end{split}
    \tag{9}
    \label{eq:generator_loss}
\end{equation}  
\begin{equation}
    \mathcal{L}_{d} = \left \| 1 - D(x) \right \| _{2}^{2}+\left \|D(\hat{x} ) \right \| _{2}^{2}
    \tag{10}
    \label{5}
\end{equation}

\noindent where $x$ and $\hat{x}$ denote the target and reconstructed speech, respectively, $\mathcal{M}(\cdot)$ is the mel-spectrogram transform. $D(\cdot)$ denotes the discriminator output and $D^{l}(\cdot)$ represents the feature map of the $l$-th discriminator layer. The sg$[\cdot]$ operator stands for the stopgradient operation. $\mathbf{z}_e$ is the output of the quantizer, and $\mathbf{e}_k$ is the codebook vector. During training, the mel-spectrograms are computed with multiple window lengths of [32, 64, 128, 256, 512, 1024, 2048] and a fixed hop length set to window length / 4. Meanwhile, different mel bin sizes of [5, 10, 20, 40, 80, 160, 320] are employed. $\lambda_{rec}, \lambda_{g}, \lambda_{feat}, \lambda_{code}$ and $\lambda_{c}$ are the empirical weights. 

\section{Experimental Settings}
\label{sec:exp}

\vspace{-1.0em}

\subsection{Training Data Preparation}
\label{ssec:data}

\vspace{-1.0em}

All training data follow the cleaning and preprocessing procedures defined in the baseline of the Challenge.\footnote{\url{https://github.com/cisco-open/lrac_data_generation}} The sampling rate of all the data is 24 kHz. Considering the attention module in the codec model and following WavTokenizer \cite{ji2024wavtokenizer}, we extract 3-second speech segments for training VoCodec with no data augmentation strategies employed.

During the training process of UL-UNAS and the final cascade system, each speech sample is mixed with background noise with 50\% probability, with the signal-to-noise ratio (SNR) uniformly distributed between -5 dB and 30 dB. Reverberation is introduced for the remaining 50\% probability, and the final training target is clean speech signals with early reverberation.

\vspace{-1.0em}

\subsection{Implementation Details}
\label{ssec:detail}

\vspace{-1.0em}

STFT is computed using a square root Hanning window of a length of 30 ms, a hop length of 10 ms, and an FFT length of 720, resulting in a buffering latency of 10 ms and an algorithmic latency of 20 ms due to the inverse STFT. 

The frequency dimension of the input feature $Z_{in}$ is first reduced from 722 to 192. Both the encoder and decoder stack 6 ConvNeXt blocks, with an intermediate dimension of 216 in the encoder and a kernel size of 7 for the depth-wise convolutional layers. For the attention module, we stack 4 ResNet blocks using a kernel size of 3 and set the group number to 2 in the decoder. Our final VoCodec comprises 3.47M parameters and has a computational complexity of 349.29M MACs/s \footnote{The computational complexity is calculated by ptflops: \url{https://github.com/tel-0s/ptflops}.} and a latency of 30 ms, with the receiver-side computation accounting for only 144.82M MACs/s. In Track 2, we adopt the same network architecture from UL-UNAS \cite{rong2025ulunasultralightweightunetsrealtime}, and scale up the number of intermediate channels to [48, 96, 108, 108, 64], which results in a computational complexity of 935.36M MACs/s. The whole system has a computational complexity of 1.28G MACs/s with 5.34M parameters. The total latency is 50 ms, of which 20 ms is attributed to the additional look-ahead in UL-UNAS.




We train UL-UNAS and VoCodec independently on 8 NVIDIA RTX 4090 GPUs. The batch size for UL-UNAS is set to 4 per GPU, while that for VoCodec is 24 per GPU. UL-UNAS and VoCodec are trained for 400 and 1000 epochs, with 1250 and 500 iterations per epoch, respectively. During training, we use the AdamW optimizer \cite{Loshchilov2017DecoupledWD} and employ a linear warmup scheduler followed by cosine annealing. 
The learning rate for UL-UNAS and VoCodec increases from $10^{-6}$ to $10^{-3}$ and $2 \times 10^{-4}$ respectively over the first 25,000 steps and then decay until 500,000 steps.
In the joint training phase, we adopt the same configuration as used in the training of VoCodec, and conduct the training process for a total of 500 epochs.

For the training of UL-UNAS, the loss weights $\lambda_1$, $\lambda_2$ and $\lambda_3$ are set to 1.0, 70 and 30, respectively. For VoCodec, the hyperparameters $\lambda_{rec}$, $\lambda_g$, $\lambda_{feat}$, $\lambda_{code}$ and $\lambda_{c}$ are set to 15.0, 1.0, 2.0, 1.0 and 0.25, respectively. In the final training phase, we employ identical loss weights to those used during the training of VoCodec.

\vspace{-1.0em}

\section{Experimental Results}
\label{ssec:result}

\vspace{-1.0em}


Evaluation on the test set is conducted using the official metrics provided by the Challenge. Scoreq-ref \cite{ragano2024scoreq}, UTMOS \cite{saeki22c_interspeech}, Audiobox AE-CE \cite{tjandra2025aes}, PESQ \cite{941023}, and Sheet-SSQA \cite{sheet, huang2024} are selected to compare the quality and naturalness of speech reconstructed by the codec. 

\begin{table}[t!]
\centering
\caption{Objective Performance Comparison on the Open Test Set of Track 1}
\label{tab3}
\small 
\resizebox{\linewidth}{!}{
\begin{tabular}{l l c c c c c c}
\toprule
Bitrate & Model & Condition & ScoreQ-ref $\downarrow$ & UTMOS $\uparrow$ & Sheet-SSQA $\uparrow$ & PESQ $\uparrow$ & \makecell{Audiobox\\AE-CE} $\uparrow$ \\
\midrule
\multirow{6}{*}{6 kbps} 
    & \multirow{3}{*}{Baseline} 
        & Clean & 0.35 & 3.23 & 3.84 & 2.67 & 5.28 \\
    & & Noisy & 0.82 & 2.76 & 3.12 & 1.81 & 4.37 \\
    & & Reverb & 1.13 & 1.32 & 2.22 & 1.18 & 3.43 \\
    \cmidrule(lr){2-8}
    & \multirow{3}{*}{VoCodec} 
        & Clean & \textbf{0.17} & \textbf{3.73} & \textbf{4.22} & \textbf{3.20} & \textbf{5.66} \\
    & & Noisy & \textbf{0.70} & \textbf{3.10} & \textbf{3.43} & \textbf{2.03} & \textbf{4.82} \\
    & & Reverb & \textbf{0.94} & \textbf{1.55} & \textbf{2.80} & \textbf{1.21} & \textbf{3.98} \\
\midrule
\multirow{6}{*}{1 kbps} 
    & \multirow{3}{*}{Baseline} 
        & Clean & 1.15 & 1.44 & 1.84 & 1.15 & 3.90 \\
    & & Noisy & 1.29 & 1.33 & 1.72 & 1.11 & 3.40 \\
    & & Reverb & 1.36 & 1.26 & 1.85 & 1.07 & 2.94 \\
    \cmidrule(lr){2-8}
    & \multirow{3}{*}{VoCodec} 
        & Clean & \textbf{0.40} & \textbf{3.24} & \textbf{3.55} & \textbf{1.95} & \textbf{5.31} \\
    & & Noisy & \textbf{0.83} & \textbf{2.67} & \textbf{2.93} & \textbf{1.56} & \textbf{4.43} \\
    & & Reverb & \textbf{1.10} & \textbf{1.48} & \textbf{2.19} & \textbf{1.17} & \textbf{3.59} \\
\bottomrule
\end{tabular}}
\end{table}

\begin{table}[t!]
\centering
\caption{Objective Performance Comparison on the Open Test Set of Track 2}
\label{tab4}
\small 
\resizebox{\linewidth}{!}{
\begin{tabular}{l l c c c c c c}
\toprule
Bitrate & Model & Condition & ScoreQ-ref $\downarrow$ & UTMOS $\uparrow$ & Sheet-SSQA $\uparrow$ & PESQ $\uparrow$ & \makecell{Audiobox\\AE-CE} $\uparrow$ \\
\midrule
\multirow{6}{*}{6 kbps} 
    & \multirow{3}{*}{Baseline} 
        & Clean & 0.43 & 2.97 & 3.55 & 2.13 & 2.97 \\
    & & Noisy & 0.75 & 2.56 & 2.92 & 1.73 & 4.60 \\
    & & Reverb & 0.92 & 1.79 & 2.67 & 1.29 & \textbf{4.25} \\
    \cmidrule(lr){2-8}
    & \multirow{3}{*}{SE + VoCodec} 
        & Clean & \textbf{0.18} & \textbf{3.74} & \textbf{4.21} & \textbf{3.06} & \textbf{5.68} \\
    & & Noisy & \textbf{0.50} & \textbf{3.26} & \textbf{3.62} & \textbf{2.19} & \textbf{5.00} \\
    & & Reverb & \textbf{0.88} & \textbf{2.02} & \textbf{2.70} & \textbf{1.38} & 4.20 \\
\midrule
\multirow{6}{*}{1 kbps} 
    & \multirow{3}{*}{Baseline} 
        & Clean & 1.01 & 1.37 & 2.07 & 1.21 & 3.96 \\
    & & Noisy & 1.15 & 1.35 & 1.95 & 1.18 & 3.70 \\
    & & Reverb & 1.12 & 1.32 & \textbf{2.43} & 1.15 & 3.55 \\
    \cmidrule(lr){2-8}
    & \multirow{3}{*}{SE + VoCodec} 
        & Clean & \textbf{0.41} & \textbf{3.21} & \textbf{3.50} & \textbf{1.92} & \textbf{5.29} \\
    & & Noisy & \textbf{0.68} & \textbf{2.81} & \textbf{3.00} & \textbf{1.63} & \textbf{4.75} \\
    & & Reverb & \textbf{1.04} & \textbf{1.75} & 2.20 & \textbf{1.26} & \textbf{3.95} \\
\bottomrule
\end{tabular}}
\end{table}

The experimental results are summarized in Table~\ref{tab3} and Table~\ref{tab4}. It can be seen that our two systems outperform the official baseline models\footnote{\url{https://github.com/cisco-open/espnet/tree/master/egs2/lrac}} across all metrics, particularly on the clean and noisy test sets. Specifically, on Track 1 we observe that even though no noise or reverberation is added during training, high-quality reconstruction of mild noise and light reverberation remains achievable at a 6 kbps bitrate, which further demonstrates the strong generalization capability of the proposed VoCodec.

Furthermore, the challenge organizers conducted subjective listening evaluations on the results of our submitted systems applied to the blind test set, and the final outcomes are presented in Table~\ref{tab1} and Table~\ref{tab2}. In both tracks, compared with the baseline models, our two systems achieve higher MUSHRA \cite{lechler25_interspeech} scores on the clean speech test set. Notably, on Track 1, VoCodec at a 6 kbps bitrate attain an MUSHRA score of 89.19, demonstrating that our proposed model can still reconstruct speech with high perceptual quality even under constraints of computational resources and low bitrates. Additionally, superior DMOS scores on noisy and reverberant test sets further indicate that VoCodec inherently possesses strong generalization capability, and its cascade system with UL-UNAS exhibits certain robustness against real-world disturbances. The improvement in diagnostic rhyme test (DRT) \cite{Lechler2024CrowdsourcedMS} scores also suggests that the speech processed by our systems achieve higher intelligibility.

\vspace{-1.0em}
\section{conclusion}
\label{sec:conclusion}

\vspace{-1.0em}

In this paper, we propose VoCodec, a model capable of reconstructing high-quality speech under constraints of limited computational resources and low bitrate. Comprehensive experiments have demonstrated that VoCodec outperforms the official baseline model for transmitted speech across multiple acoustic conditions. By cascading with UL-UNAS, VoCodec also exhibits robustness against interference in real-world adverse environments. However, the distortionless transmission of real-world speech at extremely low bitrates (e.g., 1 kbps or lower) remains a challenging task. Our future work will also focus on improving the reconstruction quality of speech under diverse acoustic scenarios at such bitrates without increasing the computational complexity.

\vspace{-1.0em}





\section{ACKNOWLEDGEMENTS}

\vspace{-1.0em}

\begin{table}[t!]
\centering
\caption{Subjective Performance Comparison on the Blind Test Set of Track 1}
\label{tab1}
\resizebox{\columnwidth}{!}{%
\begin{tabular}{c cc cc cc c}
\hline
Test Type & \multicolumn{2}{c}{Clean speech} & \multicolumn{2}{c}{Real-world light} & \multicolumn{2}{c}{Simultaneous} & Intelligibility \\
& & & \multicolumn{2}{c}{noise and reverb} & \multicolumn{2}{c}{talkers} & in clean \\
\hline
Scale & \multicolumn{2}{c}{MUSHRA} & \multicolumn{2}{c}{DMOS} & \multicolumn{2}{c}{DMOS} & DRT score \\
& \multicolumn{2}{c}{[0, 100]} & \multicolumn{2}{c}{[1, 5]} & \multicolumn{2}{c}{[1, 5]} & [-100, 100] \\
\hline
Bitrate Mode & 1kbps & 6kbps & 1kbps & 6kbps & 1kbps & 6kbps & 1kbps \\
\hline
Baseline & 17.92 & 74.28 & 1.31 & 3.35 & 1.26 & 2.20 & 75.90 \\
VoCodec & \textbf{65.20} & \textbf{89.19} & \textbf{2.74} & \textbf{4.12} & \textbf{1.70} & \textbf{2.82} & \textbf{82.98} \\
\hline
\end{tabular}%
}
\end{table}

\begin{table}[t!]
\centering
\caption{Subjective Performance Comparison on the Blind Test Set of Track 2}
\label{tab2}
\resizebox{\columnwidth}{!}{%
\begin{tabular}{c cc cc cc c c}
\hline
Test Type & \multicolumn{2}{c}{Clean speech} & \multicolumn{2}{c}{Real-world} & \multicolumn{2}{c}{Real-world} & Intelligibility in & Intelligibility in \\
& & & \multicolumn{2}{c}{speech in noise} & \multicolumn{2}{c}{speech reverb} & clean & noise \\
\hline
Scale & \multicolumn{2}{c}{MUSHRA} & \multicolumn{2}{c}{MOS} & \multicolumn{2}{c}{MOS} & DRT score & DRT score \\
& \multicolumn{2}{c}{[0, 100]} & \multicolumn{2}{c}{[1, 5]} & \multicolumn{2}{c}{[1, 5]} & [-100, 100] & [-100, 100] \\
\hline
Bitrate Mode & 1kbps & 6kbps & 1kbps & 6kbps & 1kbps & 6kbps & 1kbps & 6kbps \\
\hline
Baseline & 21.16 & 60.06 & 1.30 & 2.25 & 1.39 & 2.32 & 75.86 & \textbf{68.21} \\
SE + VoCodec & \textbf{58.63} & \textbf{75.96} & \textbf{2.09} & \textbf{2.89} & \textbf{2.15} & \textbf{3.05} & \textbf{76.21} & 66.35 \\
\hline
\end{tabular}%
}
\end{table}

This work was supported by National Natural Science Foundation of China with Grant No. 12274221 and the AI \& AI for Science Project of Nanjing University.

\renewcommand{\baselinestretch}{0.95}
\footnotesize
\bibliographystyle{IEEEbib}
\bibliography{ref.bib}

@misc{lrac,
      title={Low-Resource Audio Codec (LRAC): 2025 Challenge Description}, 
      author={Kamil Wojcicki and Yusuf Ziya Isik and Laura Lechler and Mansur Yesilbursa and Ivana Balić and Wolfgang Mack and Rafał Łaganowski and Guoqing Zhang and Yossi Adi and Minje Kim and Shinji Watanabe},
      year={2025},
      eprint={2510.23312},
      archivePrefix={arXiv},
      primaryClass={cs.SD},
      url={https://arxiv.org/abs/2510.23312}
}

@inproceedings{lechler25_interspeech,
  title     = {{Crowdsourcing MUSHRA Tests in the Age of Generative Speech Technologies: A Comparative Analysis of Subjective and Objective Testing Methods}},
  author    = {Laura Lechler and Chamran Moradi and Ivana Balic},
  year      = {2025},
  booktitle = {{Interspeech 2025}},
  pages     = {3160--3164},
  doi       = {10.21437/Interspeech.2025-2138},
  issn      = {2958-1796},
}

@article{Lechler2024CrowdsourcedMS,
  title={Crowdsourced Multilingual Speech Intelligibility Testing},
  author={Laura Lechler and Kamil Wojcicki},
  journal={ICASSP 2024 - 2024 IEEE International Conference on Acoustics, Speech and Signal Processing (ICASSP)},
  year={2024},
  pages={1441-1445},
  url={https://api.semanticscholar.org/CorpusID:268569379}
}

@inproceedings{Jang2021UnivNetAN,
  title     = {UnivNet: A Neural Vocoder with Multi-Resolution Spectrogram Discriminators for High-Fidelity Waveform Generation},
  author    = {Won Jang and Dan Lim and Jaesam Yoon and Bongwan Kim and Juntae Kim},
  year      = {2021},
  booktitle = {Interspeech 2021},
  pages     = {2207--2211},
  doi       = {10.21437/Interspeech.2021-1016},
  issn      = {2958-1796},
}

@INPROCEEDINGS{Dietz2015OverviewOT,
  author={Dietz, Martin and Multrus, Markus and Eksler, Vaclav and Malenovsky, Vladimir and Norvell, Erik and Pobloth, Harald and Miao, Lei and Wang, Zhe and Laaksonen, Lasse and Vasilache, Adriana and Kamamoto, Yutaka and Kikuiri, Kei and Ragot, Stephane and Faure, Julien and Ehara, Hiroyuki and Rajendran, Vivek and Atti, Venkatraman and Sung, Hosang and Oh, Eunmi and Yuan, Hao and Zhu, Changbao},
  booktitle={2015 IEEE International Conference on Acoustics, Speech and Signal Processing (ICASSP)}, 
  title={Overview of the {EVS} codec architecture}, 
  year={2015},
  volume={},
  number={},
  pages={5698-5702},
  doi={10.1109/ICASSP.2015.7179063}}

@misc{rfc6716,
    series =    {Request for Comments},
    number =    6716,
    howpublished =  {RFC 6716},
    publisher = {RFC Editor},
    doi =       {10.17487/RFC6716},
    url =       {https://www.rfc-editor.org/info/rfc6716},
    author =    {Jean-Marc Valin and Koen Vos and Timothy B. Terriberry},
    title =     {{Definition of the Opus Audio Codec}},
    pagetotal = 326,
    year =      2012,
    month =     sep,
}

@inproceedings{2020HiFi,
 author = {Kong, Jungil and Kim, Jaehyeon and Bae, Jaekyoung},
 booktitle = {Advances in Neural Information Processing Systems},
 editor = {H. Larochelle and M. Ranzato and R. Hadsell and M.F. Balcan and H. Lin},
 pages = {17022--17033},
 publisher = {Curran Associates, Inc.},
 title = {HiFi-GAN: Generative Adversarial Networks for Efficient and High Fidelity Speech Synthesis},
 url = {https://proceedings.neurips.cc/paper_files/paper/2020/file/c5d736809766d46260d816d8dbc9eb44-Paper.pdf},
 volume = {33},
 year = {2020}
}

@inproceedings{esser2021taming,
  title={Taming transformers for high-resolution image synthesis},
  author={Esser, Patrick and Rombach, Robin and Ommer, Bjorn},
  booktitle={Proceedings of the IEEE/CVF conference on computer vision and pattern recognition},
  pages={12873--12883},
  year={2021}
}

@inproceedings{siuzdak2023vocos,
 author = {Siuzdak, Hubert},
 booktitle = {International Conference on Representation Learning},
 editor = {B. Kim and Y. Yue and S. Chaudhuri and K. Fragkiadaki and M. Khan and Y. Sun},
 pages = {25719--25733},
 title = {Vocos: Closing the gap between time-domain and Fourier-based neural vocoders for high-quality audio synthesis},
 url = {https://proceedings.iclr.cc/paper_files/paper/2024/file/6db0903efdfe9b1bbafb015c10990b78-Paper-Conference.pdf},
 volume = {2024},
 year = {2024}
}

@inproceedings{liu2022convnet,
  title={A convnet for the 2020s},
  author={Liu, Zhuang and Mao, Hanzi and Wu, Chao-Yuan and Feichtenhofer, Christoph and Darrell, Trevor and Xie, Saining},
  booktitle={Proceedings of the IEEE/CVF conference on computer vision and pattern recognition},
  pages={11976--11986},
  year={2022}
}

@inproceedings{ji2024wavtokenizer,
 author = {Ji, Shengpeng and Jiang, Ziyue and Wang, Wen and Chen, Yifu and Fang, Minghui and Zuo, Jialong and Yang, Qian and Cheng, Xize and wang, zehan and Li, Ruiqi and Zhang, Ziang and Yang, Xiaoda and Huang, Rongjie and JIANG, YIDI and Chen, Qian and Zheng, Siqi and Zhao, Zhou},
 booktitle = {International Conference on Representation Learning},
 editor = {Y. Yue and A. Garg and N. Peng and F. Sha and R. Yu},
 pages = {93809--93826},
 title = {WavTokenizer: an Efficient Acoustic Discrete Codec Tokenizer for Audio Language Modeling},
 url = {https://proceedings.iclr.cc/paper_files/paper/2025/file/ea1f5f0878d43ff4fb8bf64ef4a2326c-Paper-Conference.pdf},
 volume = {2025},
 year = {2025}
}

@inproceedings{ragano2024scoreq,
 author = {Ragano, Alessandro and Skoglund, Jan and Hines, Andrew},
 booktitle = {Advances in Neural Information Processing Systems},
 pages = {105702--105729},
 publisher = {Curran Associates, Inc.},
 title = {SCOREQ: Speech Quality Assessment with Contrastive Regression},
 url = {https://proceedings.neurips.cc/paper_files/paper/2024/file/bece7e02455a628b770e49fcfa791147-Paper-Conference.pdf},
 volume = {37},
 year = {2024}
}

@inproceedings{saeki22c_interspeech,
  title     = {{UTMOS: UTokyo-SaruLab System for VoiceMOS Challenge 2022}},
  author    = {{Takaaki Saeki and Detai Xin and Wataru Nakata and Tomoki Koriyama and Shinnosuke Takamichi and Hiroshi Saruwatari}},
  year      = {{2022}},
  booktitle = {{Interspeech 2022}},
  pages     = {{4521--4525}},
  doi       = {{10.21437/Interspeech.2022-439}},
  issn      = {{2958-1796}},
}

@article{tjandra2025aes,
    title={Meta Audiobox Aesthetics: Unified Automatic Quality Assessment for Speech, Music, and Sound},
    author={Andros Tjandra and Yi-Chiao Wu and Baishan Guo and John Hoffman and Brian Ellis and Apoorv Vyas and Bowen Shi and Sanyuan Chen and Matt Le and Nick Zacharov and Carleigh Wood and Ann Lee and Wei-Ning Hsu},
    year={2025},
    url={https://arxiv.org/abs/2502.05139}
}

@INPROCEEDINGS{941023,
  author={Rix, A.W. and Beerends, J.G. and Hollier, M.P. and Hekstra, A.P.},
  booktitle={2001 IEEE International Conference on Acoustics, Speech, and Signal Processing. Proceedings (Cat. No.01CH37221)}, 
  title={Perceptual evaluation of speech quality (PESQ)-a new method for speech quality assessment of telephone networks and codecs}, 
  year={2001},
  volume={2},
  number={},
  pages={749-752 vol.2},
  doi={10.1109/ICASSP.2001.941023}}

@inproceedings{sheet,
  title     = {{SHEET: A Multi-purpose Open-source Speech Human Evaluation Estimation Toolkit}},
  author    = {Wen-Chin Huang and Erica Cooper and Tomoki Toda},
  year      = {2025},
  booktitle = {{Proc. Interspeech}},
  pages     = {2355--2359},
}

@article{huang2024,
      title={MOS-Bench: Benchmarking Generalization Abilities of Subjective Speech Quality Assessment Models}, 
      author={Wen-Chin Huang and Erica Cooper and Tomoki Toda},
      year={2024},
      eprint={2411.03715},
      archivePrefix={arXiv},
      primaryClass={cs.SD},
      url={https://arxiv.org/abs/2411.03715}, 
}

@article{DBLP:journals/corr/abs-2502-06490,
  publtype={informal},
  author={Yiwei Guo and Zhihan Li and Hankun Wang and Bohan Li and Chongtian Shao and Hanglei Zhang and Chenpeng Du and Xie Chen and Shujie Liu and Kai Yu},
  title={Recent Advances in Discrete Speech Tokens: A Review},
  year={2025},
  month={February},
  cdate={1738368000000},
  journal={CoRR},
  volume={abs/2502.06490},
  url={https://doi.org/10.48550/arXiv.2502.06490}
}

@article{xin2024bigcodec,
  title={BigCodec: Pushing the Limits of Low-Bitrate Neural Speech Codec},
  author={Xin, Detai and Tan, Xu and Takamichi, Shinnosuke and Saruwatari, Hiroshi},
  journal={arXiv preprint arXiv:2409.05377},
  year={2024}
}

@inproceedings{kumar2023highfidelity,
 author = {Kumar, Rithesh and Seetharaman, Prem and Luebs, Alejandro and Kumar, Ishaan and Kumar, Kundan},
 booktitle = {Advances in Neural Information Processing Systems},
 editor = {A. Oh and T. Naumann and A. Globerson and K. Saenko and M. Hardt and S. Levine},
 pages = {27980--27993},
 publisher = {Curran Associates, Inc.},
 title = {High-Fidelity Audio Compression with Improved RVQGAN},
 url = {https://proceedings.neurips.cc/paper_files/paper/2023/file/58d0e78cf042af5876e12661087bea12-Paper-Conference.pdf},
 volume = {36},
 year = {2023}
}

@ARTICLE{soundstream,
  author={Zeghidour, Neil and Luebs, Alejandro and Omran, Ahmed and Skoglund, Jan and Tagliasacchi, Marco},
  journal={IEEE/ACM Transactions on Audio, Speech, and Language Processing}, 
  title={SoundStream: An End-to-End Neural Audio Codec}, 
  year={2022},
  volume={30},
  number={},
  pages={495-507},
  doi={10.1109/TASLP.2021.3129994}}

@article{
defossez2022highfi,
title={High Fidelity Neural Audio Compression},
author={Alexandre D{\'e}fossez and Jade Copet and Gabriel Synnaeve and Yossi Adi},
journal={Transactions on Machine Learning Research},
issn={2835-8856},
year={2023},
url={https://openreview.net/forum?id=ivCd8z8zR2},
note={Featured Certification, Reproducibility Certification}
}

@INPROCEEDINGS{7780459,
  author={He, Kaiming and Zhang, Xiangyu and Ren, Shaoqing and Sun, Jian},
  booktitle={2016 IEEE Conference on Computer Vision and Pattern Recognition (CVPR)}, 
  title={Deep Residual Learning for Image Recognition}, 
  year={2016},
  volume={},
  number={},
  pages={770-778},
  doi={10.1109/CVPR.2016.90}}

@inproceedings{NIPS2017_3f5ee243,
 author = {Vaswani, Ashish and Shazeer, Noam and Parmar, Niki and Uszkoreit, Jakob and Jones, Llion and Gomez, Aidan N and Kaiser, \L ukasz and Polosukhin, Illia},
 booktitle = {Advances in Neural Information Processing Systems},
 editor = {I. Guyon and U. Von Luxburg and S. Bengio and H. Wallach and R. Fergus and S. Vishwanathan and R. Garnett},
 pages = {},
 publisher = {Curran Associates, Inc.},
 title = {Attention is All you Need},
 url = {https://proceedings.neurips.cc/paper_files/paper/2017/file/3f5ee243547dee91fbd053c1c4a845aa-Paper.pdf},
 volume = {30},
 year = {2017}
}

@INPROCEEDINGS{1171604,
  author={Biing-Hwang Juang and Gray, A.},
  booktitle={ICASSP '82. IEEE International Conference on Acoustics, Speech, and Signal Processing}, 
  title={Multiple stage vector quantization for speech coding}, 
  year={1982},
  volume={7},
  number={},
  pages={597-600},
  doi={10.1109/ICASSP.1982.1171604}}

@inproceedings{kumar2020melgan,
 author = {Kumar, Kundan and Kumar, Rithesh and de Boissiere, Thibault and Gestin, Lucas and Teoh, Wei Zhen and Sotelo, Jose and de Br\'{e}bisson, Alexandre and Bengio, Yoshua and Courville, Aaron C},
 booktitle = {Advances in Neural Information Processing Systems},
 editor = {H. Wallach and H. Larochelle and A. Beygelzimer and F. d\textquotesingle Alch\'{e}-Buc and E. Fox and R. Garnett},
 pages = {},
 publisher = {Curran Associates, Inc.},
 title = {MelGAN: Generative Adversarial Networks for Conditional Waveform Synthesis},
 url = {https://proceedings.neurips.cc/paper_files/paper/2019/file/6804c9bca0a615bdb9374d00a9fcba59-Paper.pdf},
 volume = {32},
 year = {2019}
}

@inproceedings{NIPS2017_7a98af17,
 author = {van den Oord, Aaron and Vinyals, Oriol and kavukcuoglu, koray},
 booktitle = {Advances in Neural Information Processing Systems},
 editor = {I. Guyon and U. Von Luxburg and S. Bengio and H. Wallach and R. Fergus and S. Vishwanathan and R. Garnett},
 pages = {},
 publisher = {Curran Associates, Inc.},
 title = {Neural Discrete Representation Learning},
 url = {https://proceedings.neurips.cc/paper_files/paper/2017/file/7a98af17e63a0ac09ce2e96d03992fbc-Paper.pdf},
 volume = {30},
 year = {2017}
}

@inproceedings{Loshchilov2017DecoupledWD,
  title={Decoupled Weight Decay Regularization},
  author={Ilya Loshchilov and Frank Hutter},
  booktitle={International Conference on Learning Representations},
  year={2017},
  url={https://api.semanticscholar.org/CorpusID:53592270}
}

@misc{rong2025ulunasultralightweightunetsrealtime,
      title={UL-UNAS: Ultra-Lightweight U-Nets for Real-Time Speech Enhancement via Network Architecture Search}, 
      author={Xiaobin Rong and Dahan Wang and Yuxiang Hu and Changbao Zhu and Kai Chen and Jing Lu},
      year={2025},
      eprint={2503.00340},
      archivePrefix={arXiv},
      primaryClass={eess.AS},
      url={https://arxiv.org/abs/2503.00340}, 
}

@INPROCEEDINGS{8683855,
  author={Roux, Jonathan Le and Wisdom, Scott and Erdogan, Hakan and Hershey, John R.},
  booktitle={ICASSP 2019 - 2019 IEEE International Conference on Acoustics, Speech and Signal Processing (ICASSP)}, 
  title={SDR – Half-baked or Well Done?}, 
  year={2019},
  volume={},
  number={},
  pages={626-630},
  doi={10.1109/ICASSP.2019.8683855}}

\end{document}